\documentclass[sigconf, nonacm]{acmart}

\usepackage{algorithmic}
\usepackage{graphicx}
\usepackage{textcomp}
\usepackage{subcaption}
\usepackage{soul}
\usepackage{xcolor}
\usepackage{xspace}
\usepackage{booktabs}
\usepackage{enumitem}
\def\BibTeX{{\rm B\kern-.05em{\sc i\kern-.025em b}\kern-.08em
    T\kern-.1667em\lower.7ex\hbox{E}\kern-.125emX}}

\usepackage[most]{tcolorbox}
\newtcolorbox{myboxi}[1][]{
	breakable,
	title=#1,
	colback=gray!10!white,
	colbacktitle=white,
	coltitle=black,
	fonttitle=\bfseries,
	bottomrule=0.5pt,
	toprule=0.5pt,
	leftrule=0.5pt,
	rightrule=0.5pt,
	titlerule=0pt,
	colframe=black,
	boxsep=1pt,left=2pt,right=2pt,top=2pt,bottom=2pt
}

\usepackage{makecell}

\newcommand{\system}{\textsc{NOMA}\xspace}

\newcounter{ZoiNOC}
\stepcounter{ZoiNOC}
\newcounter{IoanaNOC}
\stepcounter{IoanaNOC}
\newcommand{\zoi}[1]{\textcolor{magenta}{\small \bf [Zoi\#\arabic{ZoiNOC}\stepcounter{ZoiNOC}: #1]}}

\newcommand{\myparagraph}[1]{\vspace{0.1cm}\noindent\textbf{{#1}.~}}

\newcounter{enum}
\newenvironment{packed_enum}{
	\begin{list}{\textbf{(\arabic{enum})}}{
			\setlength{\itemsep}{0pt}
			\setlength{\parskip}{0pt}
			\setlength{\labelwidth}{-5 pt}
			\setlength{\leftmargin}{0 pt}
			\setlength{\itemindent}{0pt}
			\setlength{\topsep}{0pt}
			\usecounter{enum}}
	}{\end{list}}

\newcommand{\cmark}{{\color{green!55!black}\checkmark}}
\newcommand{\xmark}{{\color{red!70!black}$\times$}}

\newcommand\vldbdoi{XX.XX/XXX.XX}

\newcommand{\add}[1]{{\textcolor{black}{{#1}}}}

\usepackage{xcolor}
\newcommand{\revnote}[1]{\marginpar{\bfseries\textcolor{red}{} } }
\newcommand{\revtag}[1]{\textcolor{red}{\textbf{}}}

\newcommand{\rref}[1]{\hyperlink{rev:#1}{#1}}

\begin{document}

\title{Rethinking Query Optimization for Multi-Agent Systems [Vision]}

\author{Zoi Kaoudi}
\orcid{0000-0003-4520-5360}
\affiliation{%
  \institution{IT University of Copenhagen}
}
\email{zoka@itu.dk}

\author{Ioana Giurgiu}
\orcid{0000-0001-7434-7873}
\affiliation{%
  \institution{IBM Research Europe}
}
\email{igi@zurich.ibm.com }

\begin{abstract}
	The proliferation of large language models (LLMs) has accelerated the adoption of
	agent-based data pipelines. Yet current approaches remain ad hoc,
	relying on fixed structures, predefined LLMs, and single execution engines,
	without systematic optimization across heterogeneous data sources and engines.
	This paper presents \system, a query optimization framework for multi-agent data
	pipelines. We argue that optimizing agentic pipelines is a fundamentally different
	query optimization problem, with central challenges: (i)~a multi-dimensional
	search and objective space, where topology, model, and engine choices must be
	optimized jointly across latency, cost, and accuracy; (ii)~a variable pipeline
	topology; (iii)~the co-existence of diverse data models, leaving no common
	operator algebra; and (iv)~the significant cost of executing these pipelines. Our
	controlled experiment over a real-world 10-agent pipeline reveals extreme variance
	(153$\times$ cost, 5$\times$ latency, 25\% quality) and that optimal plans are heterogeneous configurations no user would construct manually.
	Our analysis of real deployments confirms these inefficiencies are systematic. We present \system{} as an integrated optimization
	loop in which plan generation, cost estimation, runtime refinement, and semantic
	caching reinforce one another across executions, setting a
	community-wide research agenda on query optimization for multi-agent systems.

\end{abstract}

\maketitle


\section{Introduction}


The rapid advancement of large language models (LLMs), together with the emergence of tool-calling interfaces and agent orchestration frameworks, has led to the rise of multi-agent data pipelines: compositions of agents that reason, invoke tools, interact with heterogeneous data sources, and collaborate to execute complex data-driven tasks. These pipelines are increasingly used in real-world applications, ranging from customer support analytics to financial monitoring and supply-chain intelligence.


\noindent\add{\textbf{Data systems for agents.}} Despite the growing adoption of such pipelines, current approaches to constructing multi-agent pipelines remain fundamentally ad hoc. Developers rely on fixed pipeline structures, pre-selected models, and single execution environments, resulting in pipelines that are often inefficient, brittle, and unable to adapt to evolving workload requirements.
In these settings, the primary unit of computation is no longer the operator but the agent, and the primary artifact is not a fixed query plan but a dynamically structured pipeline of interacting agents. This shift calls for a new direction: \emph{data systems for agents}. We must design systems where agent pipelines themselves are the object of optimization. As a result, decisions about pipeline topology, model selection, and execution engines can no longer be made independently or ahead of time: They must be jointly optimized under competing objectives such as latency, cost, and accuracy.

\noindent\add{\textbf{Challenges.}} This perspective is analogous to earlier “database for machine learning” efforts, which applied database principles to optimize ML pipelines~\cite{DB4ML}. However, multi-agent pipelines introduce new challenges. First, the optimization space is inherently multi-dimensional, requiring joint reasoning over topology, models, and engines under competing objectives. Second, pipeline topology is itself a decision variable, meaning that the optimizer must determine not only how to execute a pipeline but also how to structure it. Third, the coexistence of structured data, unstructured text, embeddings, and streaming data eliminates any common operator algebra, preventing classical pruning techniques. Finally, execution is both expensive and stochastic, as LLM calls incur significant monetary cost and variability in output quality, making naive exploration of the search space infeasible.

\noindent\add{\textbf{\system.}} In this paper, we present \system, a query optimization framework designed for multi-agent data pipelines. NOMA treats agents as first-class units of computation and extends classical optimization principles, such as cost modeling, plan generation, and re-optimization, to a setting where plans are heterogeneous, stochastic, and structurally dynamic. The optimizer produces Pareto-optimal pipelines that balance latency, cost, and quality across heterogeneous execution environments. Our contributions include:
\begin{packed_enum}
	\item We quantify the cost of suboptimal choices on a real-world pipeline, showing 153× cost variance, 5× latency variance, and a 25\% quality gap across the plan space 
	\revnote{\rref{R2.D1}}
	\add{under a fixed topology and show that varying topology reaches plans beyond it} (Section 2).
	\item We analyze multi-agent pipelines stemming from real deployments, showing that current practices leave substantial optimization potential unrealized (Section 3).
	\item We formalize the novel problem of optimizing multi-agent data pipelines (Section~\ref{sec:problem}).
	\item We present NOMA, a vision for a query optimization framework for multi-agent pipelines, organized around four research challenges: variable multi-dimensional search, heterogeneous cost estimation, continuous optimization under stochastic execution, and semantic caching of pipeline artifacts (Section 5).
\end{packed_enum}

\noindent\add{\noindent\textbf{Related work.}} \add{Existing research has largely focused on three lines of work. The first incorporates LLMs as user-defined functions within traditional query processing pipelines~\cite{patel2025lotus, FlockMTL, ThalamusDB, DocETL, palimpzest}, treating each LLM call as an operator within an otherwise fixed data system. The second targets model routing across LLMs, learning to dispatch each call to the most cost-effective model under quality constraints~\cite{FrugalGPT, routeLLM}. The third focuses on serving efficiency and orchestration on top of agent frameworks, optimizing scheduling, resource use, and execution order across LLM calls~\cite{dang, Teola, Murakkab, wadlom2026helium}. In all three, optimization targets individual operators or calls, making them cheaper, faster, or more accurate, while the overall pipeline structure remains fixed and grounded in a predefined data model and execution engine. A recent perspective~\cite{zhou2025towards}  documents redundancy and inefficiency in agentic pipelines but leaves the joint optimization across our suggested dimensions unaddressed. Table~\ref{tab:positioning} summarizes these distinctions.}

\begin{table}[t]
	\centering
	\caption{\add{Positioning of NOMA against prior work.}}
	\label{tab:positioning}
 \footnotesize                    
\setlength{\tabcolsep}{2pt}      
	\begin{tabular}{@{}lcccc@{}}
		\toprule
		\textbf{Dimension} & \makecell{\textbf{Model}\\\textbf{routing}\\\cite{FrugalGPT,routeLLM}} & \makecell{\textbf{Agent}\\\textbf{orchestration}\\\cite{Murakkab,dang,Teola,wadlom2026helium}} & \makecell{\textbf{LLM-in-DB}\\\cite{palimpzest,patel2025lotus,DocETL,FlockMTL,ThalamusDB}} & \textbf{NOMA} \\
		\midrule
		Topology a variable     & \xmark        & \cmark    & \xmark         & \cmark      \\
		Model assignment        & \cmark        & \xmark    & per-op         & \cmark      \\
		Query engine selection   & \xmark        & \xmark    & single engine  & \cmark      \\
		Objective               & cost--quality & single    & partial        & \textbf{Pareto}      \\
		Scope                   & per call      & pipeline  & operator       & \textbf{pipeline}    \\
		Runtime adaptation      & \xmark        & limited   & \xmark         & \textbf{continuous}  \\
		Cross-pipeline reuse    & \xmark        & partial\textsuperscript{*}      & \xmark   & \textbf{semantic}    \\
		\bottomrule
		\\[-6pt]
		\multicolumn{5}{@{}l}{\footnotesize\textsuperscript{*}Helium~\cite{wadlom2026helium} caches deterministic operator outputs across batches.}
	\end{tabular}
		\vspace{-0.3cm}
\end{table}

Our vision is deliberately ambitious. We require a paradigm shift: from optimizing queries over fixed plans to continuously synthesizing and adapting pipelines. Optimization becomes a persistent system function rather than a pre-execution phase. We aim to redefine query optimization for multi-agent systems, establish the foundations for a new class of query optimizers that move beyond static pipelines toward adaptive and cost-aware execution, and identify key research questions that the community must address.


\label{sec:intro}

\section{The Cost of Suboptimal Choices}
\label{sec:cost}

\begin{figure*}[t]
	\centering
	\includegraphics[width=0.9\linewidth]{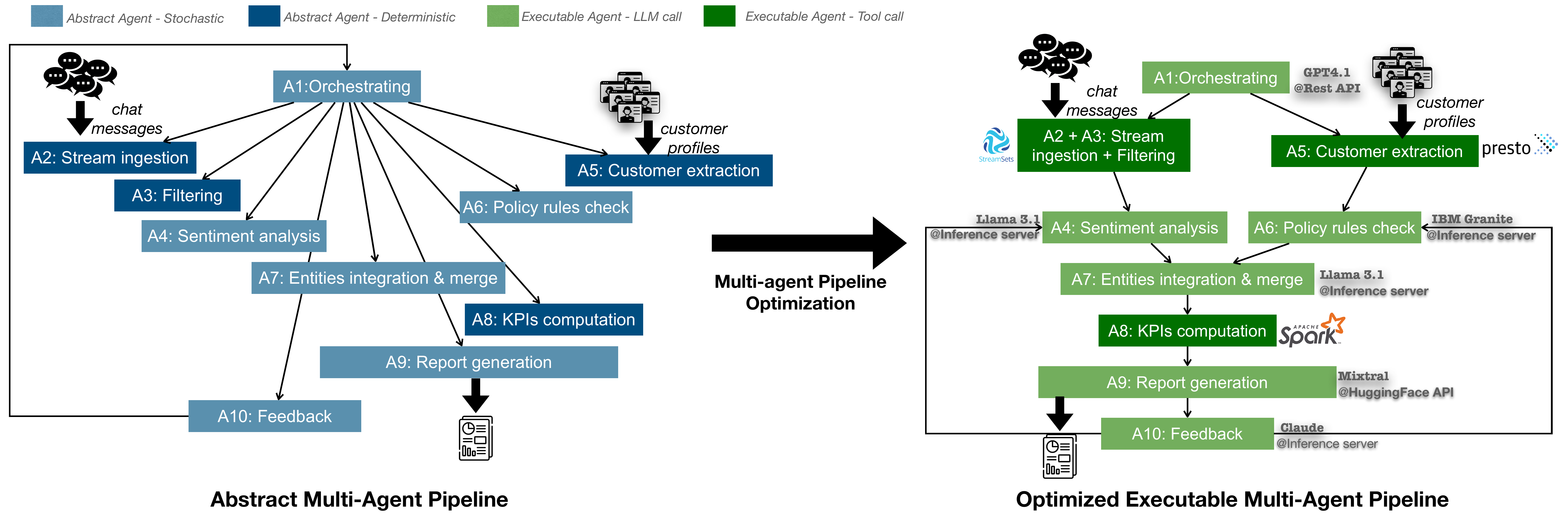}
	\caption{\add{Initial multi-agent pipeline (left) and optimized multi-agent pipeline (right).}}
	\label{fig:example}
\end{figure*}

We begin by establishing that the optimization problem is both severe and unavoidable. Using a real-world multi-agent pipeline, we show that the space of possible executions exhibits extreme variance across cost, latency, and quality, and that optimal configurations cannot be constructed manually.

\subsection{Real-world Multi-agent Pipeline}
\label{sec:example}

Consider the following real-world request from our customers:

\smallskip
\noindent\emph{Real-time Customer Support Reporting: Find all urgent cases reported by customers from live chats in the last 24 hours. Generate a detailed report and flag cases that need priority handling.}
\smallskip

This pipeline decomposes into ten agents (left part of Figure~\ref{fig:example}), combining structured data (customer profiles from a data lake), unstructured data (sentiment analysis and urgency detection), and streaming data (real-time chat messages). 
\revnote{R4.W2}
\revnote{\rref{R4.D2}}
\add{We distinguish between two classes of agents: \emph{stochastic}, whose outputs may vary across executions (light blue nodes), and \emph{deterministic}, which return the same output for the same input (dark blue nodes).}

The potential for optimization is substantial along three dimensions: how the pipeline should be structured (topology), which models should be used (model selection), and which engines should execute each task (engine selection). 
\revnote{\rref{R2.D1}}
\add{We take a first step towards quantifying this potential with a controlled experiment.}

\subsection{Experimental Setup}
\label{subsec:setup}

\myparagraph{Data} The pipeline processes approximately 2.8 GB across the four deterministic agents and between 8K and 80K input tokens per stochastic agent, depending on the task. 

\myparagraph{Engines} Each deterministic agent can be assigned to one of four engine configurations: Spark or Presto, each with 1 or 4 worker nodes. In our experimental setup, each agent has a \emph{home engine} where its data resides (A2, A3, and A8 on Spark; A5 on Presto). Running on a different engine family incurs a data transfer penalty proportional to data volume. 

\myparagraph{LLMs} 
Each stochastic agent can be assigned to one of three LLMs spanning the cost-quality spectrum: Llama 3.1 8B (cheap, fast, accuracy 0.74–0.82), Qwen 2.5-72B (moderate cost and accuracy 0.89–0.93), and Claude Sonnet 3.5 (most accurate at 0.93–0.96 but orders of magnitude more expensive). 
\revnote{R2.W4}
\revnote{\rref{R2.D3}}
\add{To account for error propagation, we measure pipeline-level quality as the weighted geometric mean of per-agent accuracy, which we obtain by task-specific evaluation against reference labels: classification accuracy for sentiment and urgency detection, F1 for entity extraction, and reference-based scoring for report generation. The weights reflect each agent’s contribution to the final output and are set uniformly. 
}

\myparagraph{Plan space in two dimensions} To make the experiment feasible, we \add{first} assume a fixed topology. A plan is a complete assignment of an engine to every deterministic agent and an LLM to every stochastic agent, yielding $4^{4} \times 3^{6} = 186{,}624$ plans. We uniformly sample 10,000 and evaluate each on monetary cost, end-to-end latency, and output quality.
To account for variance, we evaluate each sampled plan three times and report the average.
Since topology is held fixed, this space covers only two of the three optimization dimensions; the variance reported below is therefore a lower bound on what joint optimization over topology, models, and engines would reveal.
\revnote{\rref{R2.D1}}
\add{We show the topology's additional benefit at the end of this section.}

\subsection{Results}
\label{sec:results}

\begin{figure}[!t]
	\centering
	\includegraphics[width=0.6 \columnwidth]{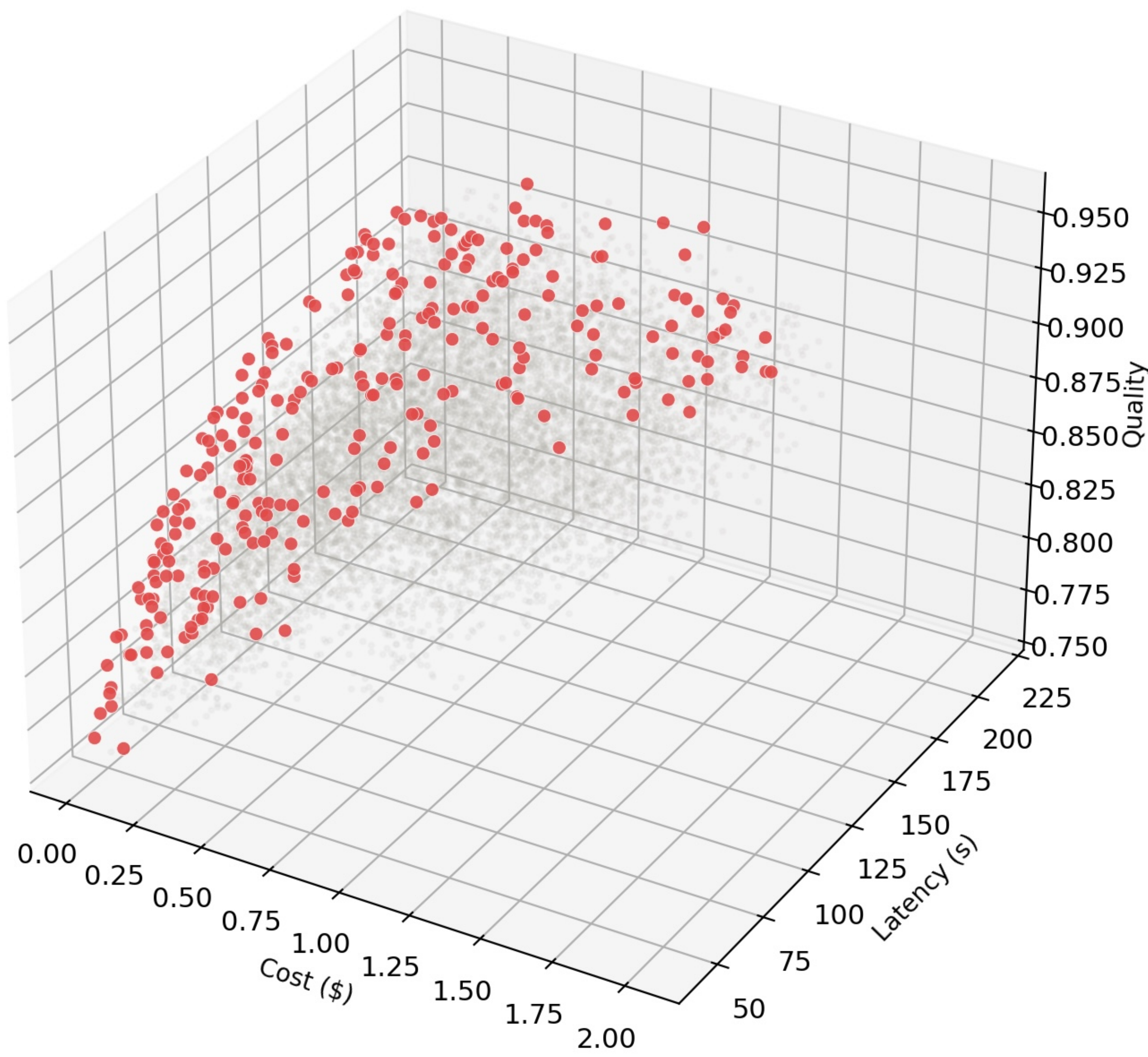}
	\caption{Pareto frontier for 10,000 randomly sampled plans.}
	\label{fig:pareto}
\end{figure}

Figure~\ref{fig:pareto} shows the three-dimensional space of cost, latency, and accuracy. We make three main observations.

\myparagraph{The plan space exhibits extreme variance} Across the 10,000 sampled plans (which explore only model and engine selection under a fixed topology), monetary cost ranges from \$0.01 to \$2.04 (a 153$\times$ spread), end-to-end latency from 42\,s to 216\,s (5.1$\times$), and quality from 0.76 to 0.95 (25\% gap). Average plans are far from optimal on any dimension, so default choices lead to poor outcomes.

\myparagraph{The Pareto frontier is rich and non-trivial} We find 266 Pareto-optimal plans, representing 2.7\% of the sampled space. The frontier spans from a cheapest plan at \$0.01 (42.5\,s latency, 0.77 quality) to a highest-quality plan at \$1.84 (125.9\,s, 0.95 quality)---a 138$\times$ cost increase for a 19\% quality improvement, while tripling the latency. These configurations are highly unintuitive: they require fine-grained coordination between model selection and execution placement that developers do not perform in practice.


\myparagraph{Optimal plans are heterogeneous} The cheapest plan (\$0.01, 42.5\,s, quality 0.77) assigns Llama 3.1 8B to all stochastic agents and places each deterministic agent on its home engine. The highest-quality plan (\$1.84, 125.9\,s, quality 0.95) mixes Claude Sonnet 3.5 and Qwen 2.5-72B across stochastic agents, deliberately using the cheaper model where Claude's advantage is marginal. A balanced plan (\$0.64, 58.9\,s, quality 0.85) cuts cost by 65\% and latency by 53\% relative to the highest-quality plan by mixing all three models. Crucially, 96\% of Pareto-optimal plans use mixed LLM assignments, a configuration that users rarely construct manually. 

\myparagraph{\add{Topology as a third optimization axis}} 
\add{The variance above is measured under a fixed topology. To be able to investigate the topology's effect as well, we modify only the structure of a balanced plan found in the previous setting. In particular, we replace the single agent A4 with a conditional refinement loop that runs the cheapest model by default and escalates to Claude only when the confidence is low. This matches the accuracy of running Claude on every case while paying the expensive model price on a small minority, so the resulting pipeline lowers both cost and latency at equal quality (cost falls from \$$0.64$ to \$$0.44$, latency from $58.9s$ to $57s$). This restructured pipeline dominates the balanced plan and lies outside the Pareto frontier of the two-dimensional space, reaching a point no model or engine assignment can reach without changing topology.
}



\section{Analysis of Multi-Agent Pipelines}
\label{sec:analysis}

Section 2 shows that simple design choices leave large portions of the plan space unexplored. We now ask whether this gap is systematic. 
\revnote{\rref{R1.O3}}
\add{We use 62 real agentic pipeline deployments spanning domains such as customer support, financial analytics, and supply-chain monitoring. We then construct and analyze 9,000 multi-agent pipelines by varying task compositions.}
Each pipeline comprises at least three tasks spanning data ingestion, schema mapping, feature engineering, analytical reasoning, and reporting. We report the following key findings.

\begin{figure}
	\begin{subfigure}{0.48\columnwidth}
		\includegraphics[width=\columnwidth]{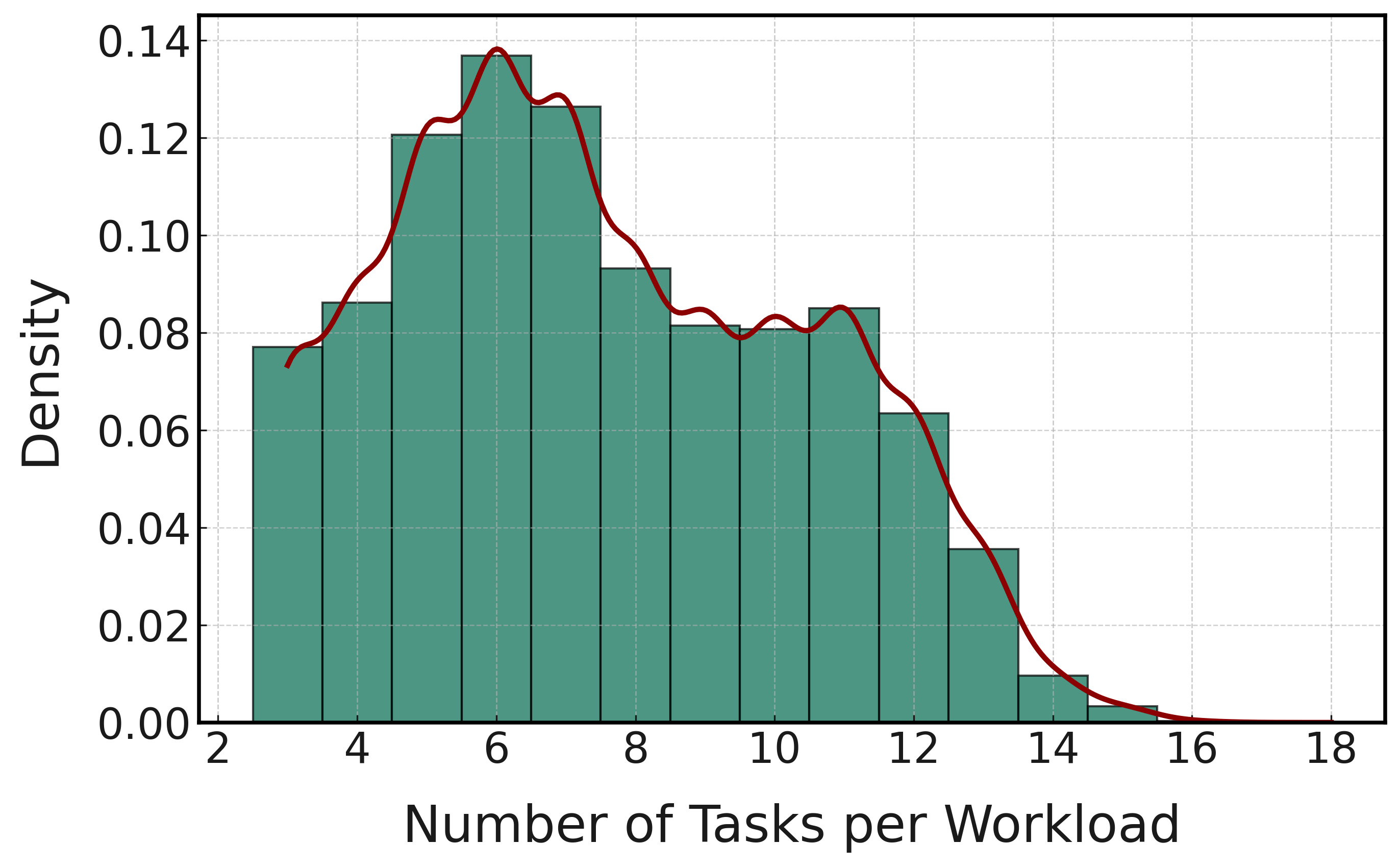}
					\vspace{-0.5cm}
		\caption{Distribution of $\#$tasks.}
		\label{fig:distribution}
	\end{subfigure}
	\hfill
	\begin{subfigure}{0.48\columnwidth}
		\includegraphics[width=\columnwidth]{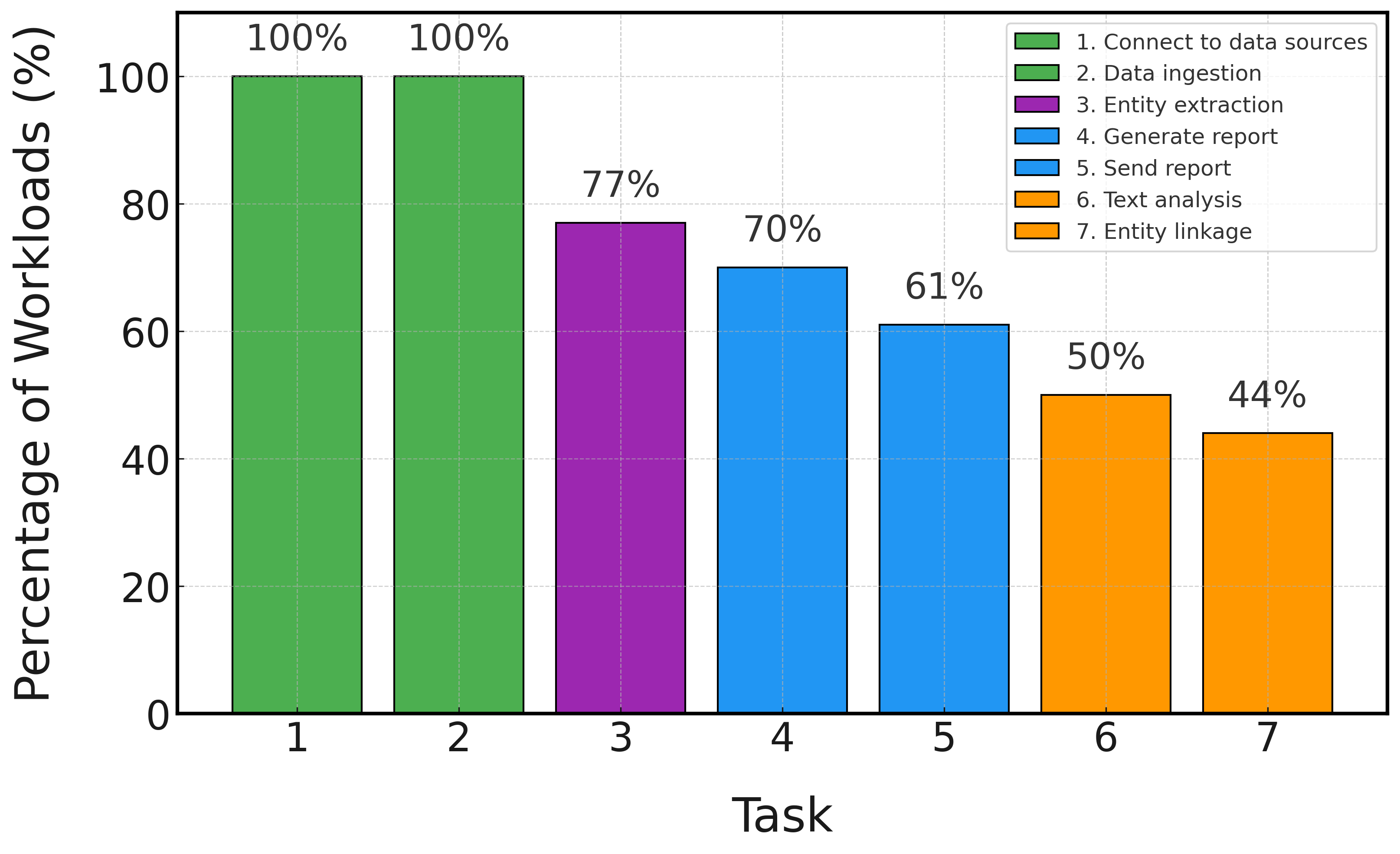}
			\vspace{-0.5cm}
		\caption{Task distribution.}
		\label{fig:redundancy}
	\end{subfigure}
	\begin{subfigure}{0.48\columnwidth}
		\includegraphics[width=\columnwidth]{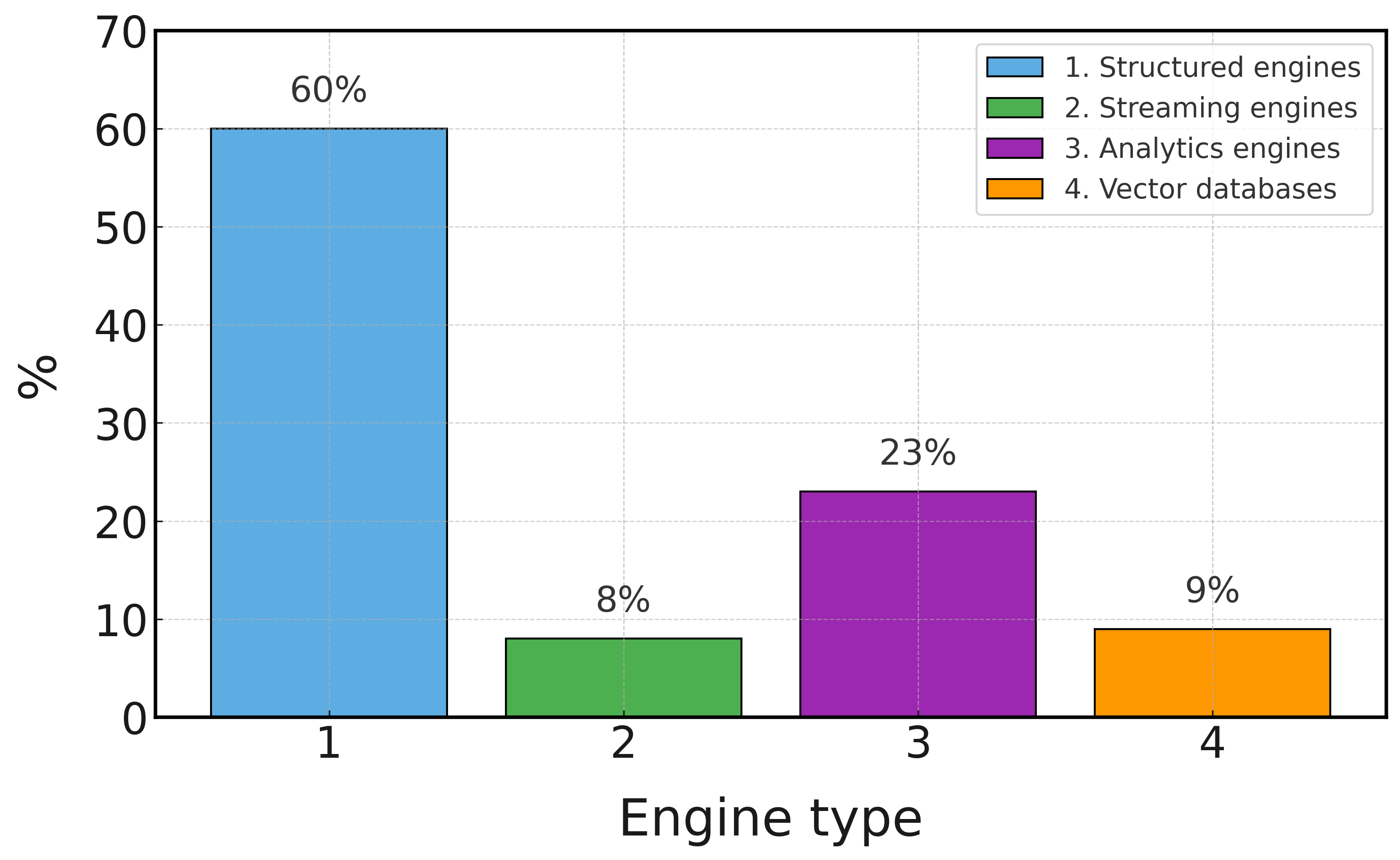}
					\vspace{-0.5cm}
		\caption{Engine selection.}
		\label{fig:engines}
	\end{subfigure}
	\hfill
	\begin{subfigure}{0.48\columnwidth}
		\includegraphics[width=\columnwidth]{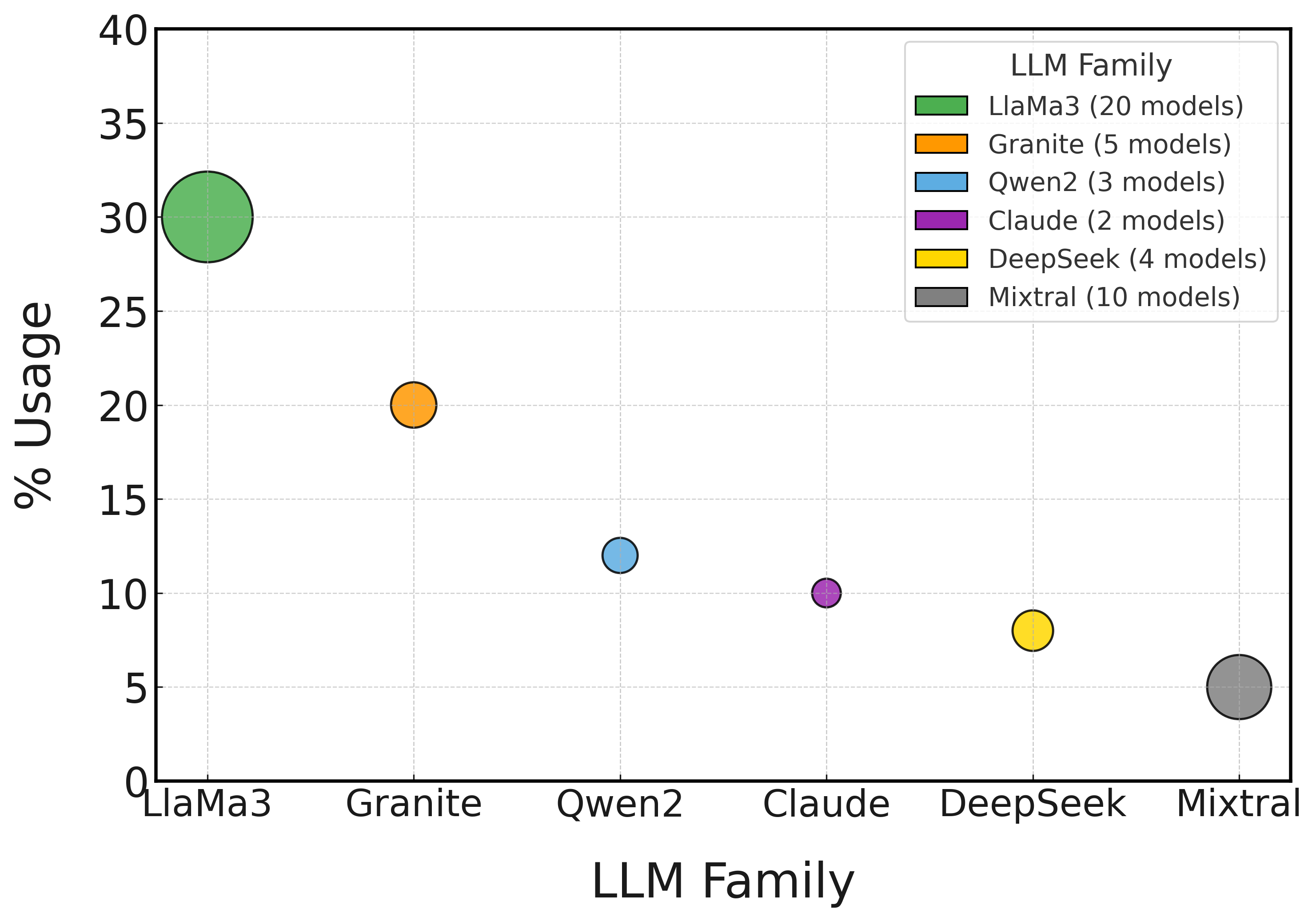}
					\vspace{-0.5cm}
		\caption{Most popular LLM families.}
		\label{fig:llms}
	\end{subfigure}
	\begin{subfigure}{\columnwidth}
		\includegraphics[width=\columnwidth]{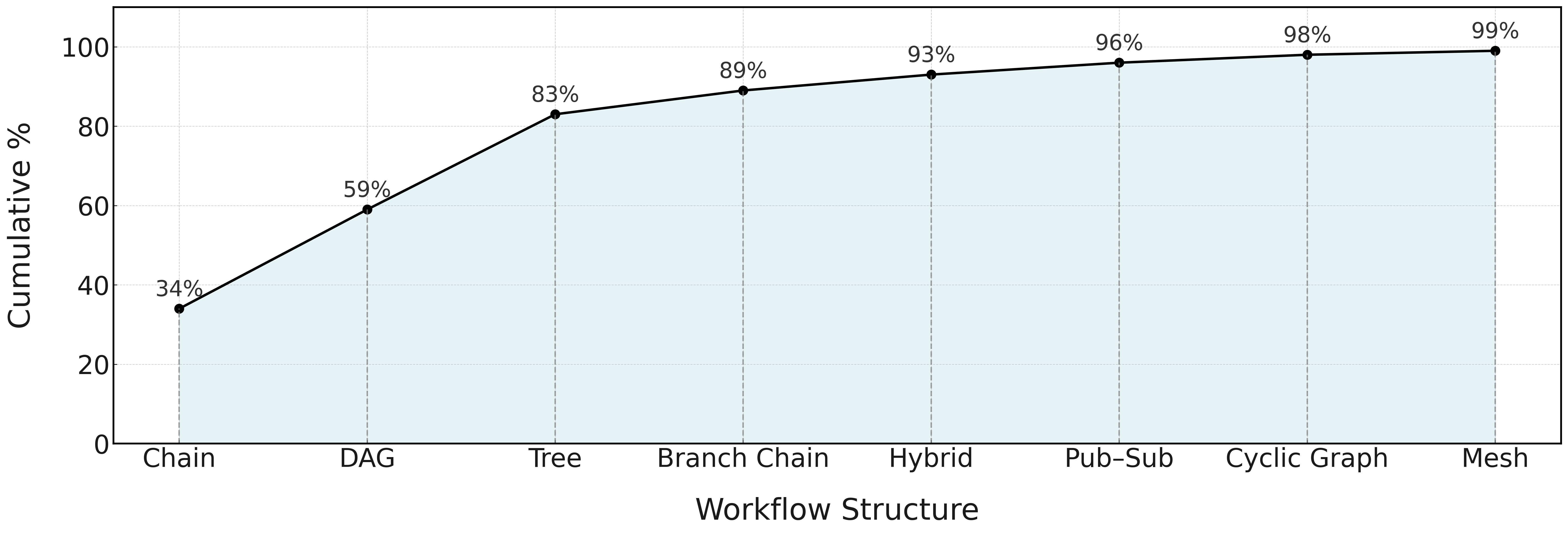}
					\vspace{-0.5cm}
		\caption{Pipelines topologies used.}
		\label{fig:architectures}
	\end{subfigure}
	\caption{Analysis of multi-agent data pipelines.\label{fig:figure1}}
\end{figure}

\myparagraph{Pipeline complexity is growing} Figure~\ref{fig:figure1}a shows the distribution of tasks per pipeline. We observe the highest density at 5--7 tasks, while pipelines with more than 10 tasks are not uncommon. As tools for automatically creating pipelines mature, we expect this distribution to shift toward even larger pipelines, much as application-generated SQL workloads have grown in complexity~\cite{sqlstorm}. This trend matters because the optimization space grows combinatorially with the number of agents.
 
\myparagraph{Model selection is driven by habit, not optimization} Across the corpus, 57\% of tasks are stochastic (Figure~\ref{fig:figure1}b), and 21\% of these use an LLM for diverse problems.
Yet when we examine which models are actually used, we observe the same limited set recurring despite the existence of a much larger model landscape (Figure~\ref{fig:figure1}d). Users strongly prefer models they have already used and tend to favor smaller models over larger ones. In other words, model selection is not driven by task-specific cost-quality trade-offs, but by familiarity and convenience. This reinforces the conclusion of Section~\ref{sec:cost}, that assigning one familiar model broadly across agents leaves substantial optimization potential unrealized.
 
\myparagraph{Simple topologies dominate despite inefficiency} The most common structure is a simple chain (34\%), followed by DAGs (25\%), trees (24\%), and branching chains (6\%) as shown in Figure~\ref{fig:figure1}e. More complex structures are rare. This distribution reflects ease of development rather than efficiency: chains expose no parallelism, and none of the dominant structures naturally capture feedback loops. 

\myparagraph{Engine heterogeneity is underexploited} 43\% of tasks involve a deterministic agent requiring a query engine (Figure~\ref{fig:figure1}c). Most of these run on structured databases (60\%), while the remainder are split across streaming engines (8\%), analytics engines (23\%), and vector databases (9\%). This confirms that 
engine assignment is rarely optimized. 
 
\myparagraph{Redundancy is pervasive} Core tasks such as data ingestion, entity extraction, and report generation recur across the majority of pipelines, yet repeated computations are rarely recognized or shared, leading to unnecessary cost and latency.

%
%
%

\section{The Optimization Problem}
\label{sec:problem}

We now formalize the problem of optimizing multi-agent data pipelines. 
Given an abstract pipeline, the goal is to determine an executable pipeline by jointly selecting pipeline topology, model assignments, and execution engines under multiple objectives such as latency, cost, and accuracy. These decisions interact in complex and non-obvious ways: A single abstract pipeline can map to many structurally different executable pipelines, each with distinct performance characteristics.

Figure~\ref{fig:example} illustrates this distinction. Starting from an abstract pipeline, agents can be merged, split, reordered, or parallelized, while model and engine assignments vary across configurations. This results in a large, heterogeneous search space.
To formalize this setting, we define the following concepts.

\myparagraph{Abstract multi-agent pipeline}
At a conceptual level, a multi-agent pipeline is a logical composition of agents collaborating to accomplish a task. Each agent provides a specific capability, such as retrieval, sentiment analysis, or summarization, and interacts with other agents. Unlike relational systems, where the space of operations is fixed, the space of tasks that agents can perform is significantly broader and, in principle, unbounded. In practice, however, a system can only execute a finite set of agents. We therefore assume the existence of a global agent registry $\mathcal{R}=\{a_1, a_2, ..., a_n\}$ that defines the available agents together with their functional descriptions and expected inputs and outputs. Formally, we define an abstract multi-agent pipeline as a directed graph $G = (A, E)$, where $A\subseteq \mathcal{R}$ is the set of agents participating in the pipeline and $E \subseteq A \times A$ captures the communication between them. Each directed edge $(a_i, a_j) \in E$ denotes a dependency in which the output of agent $a_i$ serves as input to agent $a_j$. The pipeline therefore specifies \emph{what} tasks are to be performed and the information (i.e., control and/or data) flows between agents, while abstracting away from \emph{how} each agent is implemented or executed and how they are connected. We distinguish between two classes of agents: \emph{stochastic} agents, whose outputs may vary across executions and may be approximate (e.g., summarization), and \emph{deterministic} agents, which return the same output for the same input (e.g., database retrieval).

\myparagraph{Executable multi-agent pipeline}
Given an abstract pipeline $G$, an executable pipeline specifies (i)~the model assigned to each stochastic agent, (ii)~the execution engine for each agent, and (iii)~the structure that determines execution order, parallelism, and communication. Formally, an executable pipeline is a graph $G^*=(A^*, E^*)$ that may differ structurally from $G$, where $A^*$ denotes the set of agents with concrete model and engine instantiations and $E^*$ shows the communication among the instantiated agents.

\myparagraph{\add{Transformations}}\revnote{\rref{R2.D4}}\add{$G^*$ may have a different structure from $G$ because of a	structure-changing transformation. We consider four transformations: (i)~\emph{fusion} merges two adjacent agents when the first feeds only the second; (ii)~\emph{split} decomposes an agent when its sub-agents reproduce its input-output contract; 
(iii)~\emph{reorder/parallelize} repositions or parallelizes agents at any point where their inputs remain available; and \emph{conditional re-execution} adds a bounded guarded loop that re-invokes an agent while its quality estimate is low. In all cases the producer-consumer dependencies and each agent's input-output contract are preserved, so that $G^*$ stays semantically consistent with $G$.}

\myparagraph{Problem statement}
Given an abstract pipeline $G = (A, E)$ and a set of optimization objectives $O = {o_1, \dots, o_m}$ (e.g.,~$\#$tokens, monetary cost, quality, runtime, energy consumption, etc.), the goal is to find one or more executable pipelines $G^*=(A^*, E^*)$ that are Pareto-optimal with respect to $O$.


\add{Note that among the objectives in $O$, quality is distinctive: unlike monetary cost or runtime, it has no universal metric and is task dependent. 
We therefore define the quality of an executable pipeline $G^*$ as a task-specific function $q(G^*)$ which can be instantiated in different ways, e.g.,~via reference-based metrics (precision, recall, F1, exact match), model-based judges (LLM-as-judge, self-consistency), or domain-specific scores.
The weighted geometric mean of per-agent accuracy used in Section~\ref{sec:cost} is one such instantiation.
} 

This problem formulation shows why multi-agent pipeline optimization cannot be reduced to traditional query optimization. First, topology is a variable, rather than a plan with fixed operators. Second, optimization is inherently multi-objective, requiring trade-offs across latency, cost, and accuracy. 
Third, the absence of a common operator algebra prevents the use of classical pruning techniques. 
Together, these properties invalidate the core assumptions underlying traditional optimizers and require a fundamentally different approach, motivating the design of \system.

\section{Vision and Research Challenges}
\label{sec:vision}
Optimizing multi-agent pipelines requires rethinking each pillar of query optimization. 
We identify the main challenges that arise in this new setting and for each one we describe why existing approaches are insufficient, present our vision for how \system addresses it, and identify the open research questions that remain.

 \subsection{Variable Multi-Dimensional Search Space}

  The search space of a multi-agent pipeline spans three dimensions: pipeline topology, model selection, and execution engine assignment. These are highly inter-dependent: changing a model assignment changes the downstream workload that engine decisions must account for, merging two agents eliminates a model decision, while splitting an agent creates new ones. 
 The set of decision variables is thus not fixed but changes with each structural alternative, yielding a vast, variable space where exhaustive enumeration is infeasible.
 Besides the complexity of the search space, the multi-objective nature of the problem compounds this difficulty. The system must explore a Pareto frontier over different user-specified objectives. The search process must create diverse candidates that reflect different trade-offs rather than converging to a single optimal plan.
 These challenges rule out classical enumeration-based approaches, where algebraic equivalences prune the search space and cost models estimate a single cost for the candidates.
 
 \myparagraph{Our vision} We propose using generative models trained on historical pipelines to generate executable pipelines directly in the joint optimization space, bypassing plan enumeration. In \system, a \emph{multi-agent pipeline generator} produces candidates that vary across topology, model assignments, and engine configurations simultaneously, reflecting different positions on the Pareto frontier. Generative models have recently been proposed for relational query optimization~\cite{GenJoin}, however this work operates over a fixed relational plan structure and optimizes for a single objective (execution time). 
 Similarly, learned orchestrators~\cite{dang} optimize execution order for a fixed agent set under a single objective.
In contrast, NOMA's generator must produce both pipeline structure and execution decisions jointly, without a fixed plan shape, without cardinality estimates, and across a multi-objective Pareto frontier, making it a fundamentally different generation problem. 
%

\myparagraph{Open research questions} Training generative models requires examples
of good pipelines, yet obtaining them requires
executing many non-optimal pipelines: How can we bootstrap training data without exhaustive, costly
exploration? What representations capture both graph structure and agent semantics
well enough to generalize to unseen pipelines? And how can generation be made
cost-aware, so that candidates are likely near the Pareto frontier before the cost
model evaluates them, avoiding wasted evaluations on poor candidates?

\subsection{Unified Cost Models}
\label{subsec:costmodel}

Traditional cost models, including learned ones, assume a fixed plan structure and produce scalar estimates, assumptions that break in the multi-agent setting.
First, topology is a variable: the cost model must compare pipelines differing in
the number of agents, their connectivity, and their execution order. Learned
relational cost models~\cite{vldb-tutorial-2025,giurgiu2025,LCM-experimental-sigmod25}
predict cost only within a given plan shape, while cross-engine
optimizers~\cite{rheem, wayang-sigmod-record} extend this to heterogeneous engines
but still assume a fixed structure and a shared relational data model---neither of
which holds here. Second, agent cost signals are not directly comparable:
LLM agents incur token costs, stochastic latency, and variable quality, database
agents depend on data volume and indexing, and streaming engines differ again. 
Quality propagation compounds this: a cheaper, lower-quality upstream
model changes the volume and nature of downstream inputs, altering engine costs
and model requirements in ways a per-agent cost model cannot capture.

\myparagraph{Our vision} We propose building \emph{unified cost models} that map heterogeneous agent features into a shared representation and predict calibrated distributions over latency, monetary cost, and accuracy rather than scalar point estimates. Rather than a single monolithic model, we envision a family of cost models, one per objective, that share a common pipeline graph representation but specialize their output heads for the statistical properties of each objective. Distributional estimates are essential because agents are stochastic and quality propagates non-additively: a point estimate cannot capture the uncertainty introduced by upstream model choices, making it impossible to reason reliably about downstream costs. 
Conformal prediction~\cite{conformal, LCM-conformal} can provide distribution-free calibration guarantees. Online recalibration~\cite{limao-optimizer} keeps the models accurate as pricing, hardware availability, and workloads shift.
\revnote{R2.W3}
\revnote{\rref{R2.D5}}
\add{The three objectives differ significantly in how they can be learned. Monetary cost can be computed almost analytically: for stochastic agents it is a deterministic function of token volume and published per-token prices and for deterministic agents a function of data volume and resource-time at known rates; the only learned component is predicting volume before execution. Latency is largely profiled offline: per-agent latency distributions are measured per (model, engine, input-size) and composed along the pipeline, with a learned correction for contention when agents share resources. Quality is the only objective that is fully learned, label-scarce, and non-additive, and is therefore where the challenge concentrates.
\revnote{\rref{R2.D4}}
Context passing is accounted for when an execution plan is evaluated: the context an agent receives determines its input cost and latency and affects its output quality.
Critically, the cost model need not be accurate in an absolute sense: it only needs to rank candidates well enough to prune generation and stay calibrated to flag when an execution deviates from prediction, triggering the refiner.
}

\myparagraph{Open research questions} How can cost models compare configurations
across structurally different topologies? How should heterogeneous cost signals
(token usage, data volume, model accuracy) be reconciled into a unified
representation? How can distributional estimates stay calibrated when quality
propagates non-additively, so that uncertainty at one agent affects all downstream
agents?
\revnote{\rref{R2.D3}}
\revnote{R2.W4}
\revnote{R4.W1}
\revnote{\rref{R4.D1}}
\revnote{R4.R1}
\add{And how should pipeline quality be defined and estimated efficiently, given
	that it is task-dependent and lacks a directly observable label?}

\subsection{Optimization Under Stochastic Execution}
Traditional query optimization is a one-shot process: the optimizer produces a plan before execution begins and the executor carries it out without further intervention. This separation is possible because relational operators are deterministic and cost model errors manifest only as suboptimal performance, not as fundamentally changed downstream workloads. Multi-agent pipelines violate both conditions. Stochastic agents introduce non-determinism: the same input can produce different outputs on different runs, and the quality of an upstream agent's output directly affects the volume, distribution, and nature of downstream inputs. 
One-shot optimization cannot account for this because the actual workload is not known until execution begins. Existing approaches such as progressive optimization~\cite{markl2004robust} and adaptive query processing~\cite{deshpande2007aqp} do not address this setting:
they trigger re-optimization reactively when observed cardinalities deviate from estimates and re-planning is cheap relative to execution. In our setting, revision must be continuous and re-planning is expensive enough that the system must reason about whether the expected benefit of revision justifies its cost.


\myparagraph{Our vision} We argue that multi-agent pipeline optimization should be continuous. In \system, a \emph{continuous pipeline refiner} is a first-class component that participates in execution from start to finish. It uses a Bayesian optimization process where each agent completion yields an observation that updates cost model priors and revises downstream decisions accordingly. The refiner then applies an acquisition function to decide whether the expected gain from re-planning justifies its cost.
BayesQO~\cite{offline-optimization} has shown the effectiveness of Bayesian optimization for navigating a plan space in relational query optimization. In contrast, \system's refiner must handle multi-objective optimization over a variable pipeline topology rather than a single objective over a fixed query structure.
The refiner is tightly coupled with the cost models of Section~\ref{subsec:costmodel}.

\myparagraph{Open research questions} How should the scope of revision be determined after each agent completion: revising only the next decision is cheap but myopic, while revising the full remaining pipeline is expensive but globally optimal? How can the system decide when revision pays off, given that planning cost is no longer negligible? How can stability be ensured under frequent revisions?

\subsection{Semantic Caching Across Pipelines}

Multi-agent workloads exhibit significant redundancy: data ingestion, entity extraction, summarization, and report generation recur across pipelines and often over overlapping inputs (see Section~\ref{sec:analysis}). In a setting where each LLM execution incurs non-trivial monetary cost, exploiting this redundancy through caching is not an optional enhancement but a core cost reduction mechanism. Traditional caching relies on syntactic equivalence — a cached result can be reused when queries match exactly or can be shown equivalent through algebraic rewriting. This assumption breaks down entirely in the multi-agent setting. Agent inputs are expressed in natural language, executions are non-deterministic, and there are no algebraic rewriting rules to establish equivalence. Two agent invocations that are semantically equivalent will not match syntactically, yet reusing a cached result may be entirely appropriate.
Semantic caching has been studied since Dar et al~\cite{Dar1996SemanticCaching} introduced predicate-based caching. 
Recent systems adapt it to LLM workloads via embedding similarity over queries~\cite{bang-2023-gptcache, Gill2025MeanCache,Yan2025ContextCache}. These cache individual query--response pairs over flat
inputs; they do not cache sub-plans or optimization decisions, reason about
structural similarity across pipelines, or transfer configurations to
structurally different pipelines---the challenges \system's cache must address.

\myparagraph{Our vision} We propose a \emph{multi-purpose cache} that goes beyond caching intermediate results to also storing optimized sub-plans and optimization decisions: model assignments and engine configurations that proved effective in prior pipelines. The cache indexes all these artifacts semantically rather than syntactically. It combines structural normalization with semantic embedding, with a lightweight coarse filter to eliminate obvious mismatches before invoking the more expensive embedding-based similarity check. 

\myparagraph{Open research questions} How should semantic equivalence be defined
for agent inputs, balancing precision, recall, and similarity-computation cost?
How can sub-plan reuse be supported across structurally different pipelines,
where similarity is semantic rather than syntactic? And how should staleness be
managed, and cached decisions transferred across pipelines?

\subsection{\system as an Integrated Optimization Loop}
\label{subsec:evaluation-plan}
The four challenges above are tightly coupled facets of a single optimization loop. 
\revnote{R1.O1}
\revnote{\rref{R1.D3}}
\add{\system's contribution is that its components are mutually dependent and must be co-designed. }
The generator proposes candidates, the cost models score them, the refiner revises decisions at runtime, and the cache closes the loop by converting execution experience into reusable knowledge. 
\add{Each of our envisioned solutions extends an existing line of work and does not necessarily constitute a standalone contribution. As a concrete agenda, we propose to address the challenges in the following order: (1) a unified cost model over a fixed topology; (2) pipeline generation built on top of it; (3) the continuous pipeline refiner; and (4) caching.}

\myparagraph{\add{Evaluation}}
\add{
We will evaluate \system{} on a representative subset of our 9,000 pipelines, against baselines spanning a developer-default ``habit''}
\revnote{R1.O2}
\revnote{\rref{R1.D4}}
\add{
configuration, per-agent model routing, orchestration-only approaches, random
sampling with Pareto filtering, and an oracle optimizer on small instances as an
upper bound. We will report end-to-end cost, latency, and quality. In addition, we will measure optimization overhead, distance to a reference Pareto frontier, and component-level contributions, such as cache hit rate and refinement gains.
We expect \system to deliver substantial reductions in monetary cost and latency at matched quality levels relative to developer-default baselines, while approaching the Pareto frontier obtained by exhaustive enumeration on small pipelines.
}


\section{Looking Ahead}
\label{sec:conclusion}

Multi-agent data pipelines represent a fundamental shift in how data-driven tasks are executed. Yet, they remain largely unoptimized. We have shown that this gap is consequential: the plan space exhibits extreme variance, current practices leave substantial optimization potential unrealized, and optimal configurations are ones no developer would construct manually. The challenges of jointly optimizing topology, model selection, and execution engines require rethinking each pillar of query optimization, from plan generation and cost estimation to runtime refinement and caching. \system is our vision for a framework that addresses these challenges as an integrated optimization loop, where each pipeline execution makes the next one cheaper and better. 

If optimizers like \system become available, multi-agent pipelines cease to be static programs and instead become continuously synthesized and refined artifacts. Users no longer define pipelines explicitly, but specify objectives and constraints, leaving the system to construct, adapt, and replace pipelines as models, data characteristics, and pricing evolve.

This shift enables fundamentally new behaviors. Optimization becomes persistent and workload-wide, with execution experience accumulated and reused across tasks. Systems can arbitrage across models, engines, and time — deciding not only how to execute but when and whether to execute under cost, latency, and quality trade-offs. The 153× cost variance we observed in Section~\ref{sec:cost} is not static: model pricing fluctuates, provider rate limits vary by time of day, and spot-instance availability shifts. A system that reasons over this temporal dimension can defer non-urgent sub-pipelines to cheaper windows or batch stochastic agents when provider load is low, extracting savings that no fixed-time optimizer can capture. Pipeline configurations themselves become self-evolving, with topology, model choices, and execution strategies continuously rewritten as the optimizer accumulates evidence about what works. 

More broadly, this direction aligns with emerging work on synthesized data systems, where database engines and query processing strategies are generated from workload requirements rather than engineered~\cite{binnig2026bespoke, trummer2026gendb}. \system extends this trajectory to multi-agent workloads: instead of optimizing queries over a fixed system, it enables pipelines, execution strategies, and system configurations to co-evolve. In this view, optimization is no longer a stage in execution but a continuous function, and agentic data systems transition from fixed pipelines to adaptive, self-optimizing infrastructures. We hope this paper serves as a call to action for the database, systems, and ML communities to take on this problem together.

\bibliographystyle{ACM-Reference-Format}
\bibliography{refs}

\end{document}